\documentclass[%
superscriptaddress,
preprint,
 amsmath,amssymb,
 aps,
]{revtex4-2}

\usepackage[english]{babel}
\usepackage{graphicx}
\usepackage{dcolumn}
\usepackage{physics}
\usepackage{bm}
\usepackage{hyperref}
\usepackage{cleveref}
\usepackage{fancyhdr}
\usepackage{romannum}
\usepackage{enumerate}
\usepackage{natbib}
\usepackage{tabularx}
\usepackage{floatrow}
\usepackage[label font=bf,labelformat=simple]{subfig}
\usepackage{caption}
\usepackage{setspace}

\newcommand{\myref}[1]{(\ref{#1})}

\graphicspath{{figure/}}

\begin{document}

\renewcommand{\thepage}{\arabic{page}}

\title{Universality of noise-induced resilience restoration in spatially-extended ecological systems}

\author{Cheng Ma}
\affiliation{Department of Physics, Applied Physics and Astronomy, Rensselaer Polytechnic Institute, \\
Troy, NY 12180, USA }
\affiliation{Network Science and Technology Center, Rensselaer Polytechnic Institute, Troy, NY 12180, USA}
\author{Gyorgy Korniss}
\affiliation{Department of Physics, Applied Physics and Astronomy, Rensselaer Polytechnic Institute, \\
Troy, NY 12180, USA }
\affiliation{Network Science and Technology Center, Rensselaer Polytechnic Institute, Troy, NY 12180, USA}
\author{Boleslaw K. Szymanski}
\affiliation{Department of Physics, Applied Physics and Astronomy, Rensselaer Polytechnic Institute, \\
Troy, NY 12180, USA }
\affiliation{Network Science and Technology Center, Rensselaer Polytechnic Institute, Troy, NY 12180, USA}
\affiliation{Department of Computer Science, Rensselaer Polytechnic Institute, Troy, NY 12180, USA}
\author{Jianxi Gao}
\affiliation{Network Science and Technology Center, Rensselaer Polytechnic Institute, Troy, NY 12180, USA}
\affiliation{Department of Computer Science, Rensselaer Polytechnic Institute, Troy, NY 12180, USA}

\begin{abstract}
Many systems may switch to an undesired state due to internal failures or external perturbations, of which critical transitions toward degraded ecosystem states are a prominent example. Resilience restoration focuses on the ability of spatially-extended systems and the required time to recover to its desired state under stochastic environmental conditions. While mean field approaches may guide recovery strategies by indicating the conditions needed to destabilize undesired states, these approaches are not accurately capturing the transition process toward the desired state of spatially-extended systems in stochastic environments.
The difficulty is rooted in the lack of mathematical tools to analyze systems with high dimensionality, nonlinearity, and stochastic effects. We bridge this gap by developing new mathematical tools that employ nucleation theory in spatially-embedded systems to advance resilience restoration. We examine our approach on systems following mutualistic dynamics and diffusion models, finding that systems may exhibit single-cluster or multi-cluster phases depending on their sizes and noise strengths, and also construct a new scaling law governing the restoration time for arbitrary system size and noise strength in two-dimensional systems. This approach is not limited to ecosystems and has applications in various dynamical systems, from biology to infrastructural systems.	
\begin{description}
\item[Keywords] Alternative stable states, Resilience, Nucleation theory, Noise-induced transitions
\end{description}
\end{abstract}

\maketitle

\section{\label{sec:intro}Introduction}

Resilience, a system's ability to retain its basic functionality when errors and failures occur, is a defining property of many complex systems \cite{cohen2000resilience,hollingResilienceStabilityEcological1973,gaoUniversalResiliencePatterns2016}. Theoretical models of resilience loss and transitions between alternative states are used to study unanticipated and drastic changes. In real-world systems, many critical transitions are observed, including catastrophic shifts in ecological systems \cite{schefferCatastrophicShiftsEcosystems2001, schefferGenericIndicatorsEcological2015}, blackouts in power grids \cite{dobsonComplexSystemsAnalysis2007}, financial crises \cite{mayEcologyBankers2008}, climate changes \cite{lentonTippingElementsEarth2008}, human depression \cite{vandeleemputCriticalSlowingEarly2014}. These abrupt shifts may arise in the presence of alternative stable states. They can be modeled as critical transitions causing the resilience loss of the desired state, after which the systems switch from functional to a dysfunctional state \cite{mayThresholdsBreakpointsEcosystems1977, feudelComplexDynamicsMultistable2008, guttalImpactNoiseBistable2007}. Ecologists are particularly concerned about sudden critical switches between alternative stable states \cite{schefferCatastrophicShiftsEcosystems2001, schefferGenericIndicatorsEcological2015}. As the system approaches the tipping point, its behavior becomes barely predictable when the gradual change of some environmental factors leads to the drastic change of the system state. Once a system loses its resilience and becomes dysfunctional, restoring the environmental conditions only to those existing before the collapse is often insufficient. Instead, environmental conditions should be recovered to the critical point where the undesired state is destabilized and then resilience restoration would occur. In this work, we study the resilience restoration in spatially-extended ecosystems, which focuses on a system's ability and required time to recover to its desired state after the transition to the undesired state. Even though various restoration methods targeted at specific systems have been proposed, stochastic perturbations receive little attention \cite{carpenterResilienceRestorationLakes1997, whiteResilienceRestorationRiparian2011, devotoUnderstandingPlanningEcological2012, lakeResistanceResilienceRestoration2013, buissonResilienceRestorationTropical2019, almoghathawiResiliencedrivenRestorationModel2019}. On the other hand, stability of stochastic dynamical systems containing only one variable has been extensively explored. Also, it has been known for a long time that noise can greatly affect the stability and resilience of the system with bistable states. In particular, it has been shown that noise can induce transitions between alternative stable states and the required time to transition has been established by computing the mean first passage time (MFPT) \cite{hollingResilienceStabilityEcological1973, yangDelayNoiseInduced2014, zengNoisesinducedRegimeShifts2015, dodoricoNoiseinducedStabilityDryland2005}. Those studies indicate that the single-variable systems can switch between alternative stable states in the presence of noise.
One may expect that noise can also induce transitions in multi-variable systems. However, one should keep in mind that such noise-induced transition occurs only when the system is close to the bifurcation point where the basin of attraction of the current stable state vanishes. If the system is far from this bifurcation point, noise may drive the system back and forth between alternative stable states, or even not be able to trigger any transition. Since environmental stochasticity is an inherent property of real-world ecosystems, we investigate resilience restoration after introducing stochastic perturbations into multi-variable systems, providing a theoretical understanding of critical transitions in spatially-extended systems subject to environmental stochasticity. 

Despite advances in understanding the macroscopic characteristics of resilience restoration, previous research has mostly focused on single-variable or low-dimensional systems, which do not account for the exceptionally large number of variables that in reality are needed to control the state of a complex system. Indeed, many real-world systems consist of numerous components connected via a complex set of weighted, often directed, interactions \cite{barabasi2016network}. The complicated interactions may lead to phenomena not arising in single-variable systems. For example, in the ecosystems, the recovery (or extinction) of a species in one location can impact the states of this species in the neighboring locations, leading to a recovery (or extinction) over the entire system \cite{sanhedrai2020reviving}. Accordingly, a full understanding of the system evolution, stability, and resilience cannot be gained without considering interactions among a sufficiently large number of components. However, hindered by the high-dimensionality of interaction topology and the nonlinear evolution dynamics, few analyses of critical transitions and resilience restoration had been done directly on high-dimensional systems consisting of a great number of participants until the effective reduction theory was recently developed by Gao \emph{et.al.} \cite{gaoUniversalResiliencePatterns2016}. This theory enables us to effectively reduce a multi-dimensional complex system to a one-dimensional system by capturing the average activities of the original system. Furthermore, Liang \emph{et.al.} \cite{liangUniversalIndicatorCritical2017} designed a universal indicator for critical transitions in complex networks and concluded that noise compensates for the structural defects of complex networks, indicating that noise may alter the critical threshold. Jiang \emph{et.al.} \cite{jiangPredictingTippingPoints2018} studied mutualistic networks through dimensional reduction and claimed that the tipping point can be predicted accurately even in the presence of noise. Tu \emph{et.al.} \cite{tu2021dimensionality} developed an analytical framework for collapsing complex $N$-dimensional networked systems into an $M+1$-dimensional manifold as a function of $M$ effective control parameters with $M \ll N$. Nonetheless, our study shows that noise eliminates the deterministic critical threshold, and the recovery of the entire system from the dysfunctional state is possible in the presence of perturbations as long as noise can trigger the transition for just one component. This scenario is very likely to occur when the system is close to the deterministic threshold where the undesired state loses stability. The farther away the system is from this point, the more difficult it is for noise to induce the transition.

The time required to recover a system is a quantity of great interest but determined by many factors, including system sizes, noise strengths, and dynamical functions. Nucleation theory provides an elegant bridge between the noise-induced transition and the spread of such transition over the entire system. The classical approach to homogeneous nucleation theory was originally developed to describe phase transformation in materials \cite{Kolmogorov_1937, Johnson_1939, avramiKineticsPhaseChange1939, avramiKineticsPhaseChange1940, lotheReconsiderationsNucleationTheory1962}, in particular in ferromagnetic \cite{Rikvold_PRE1994, Ramos_PRB1999} and ferroelectric systems \cite{Ishibashi_JPSJ1971, Duiker_PRB1990}. Recently, it was applied to invasion phenomena in spatial ecological systems \cite{gandhiNucleationRelaxationMetastability1999, kornissSpatialDynamicsInvasion2005, omalleyInvasiveAdvanceAdvantageous2006, Allstadt_EER2007}. Korniss \emph{et.al.} \cite{kornissSpatialDynamicsInvasion2005} and O'Malley \emph{et.al.} \cite{omalleyInvasiveAdvanceAdvantageous2006} studied ecological invasion in spatially-extended systems with competition. They discriminate between two fundamental modes of nucleating invasive clusters (single-cluster vs. multi-cluster) and their time-evolution and stochastic features.
A more recent study by Michaels \emph{et.al.} \cite{michaelsNucleationFrameworkTransition2020} combined nucleation theory with local-scale positive feedback and offered a novel way to understand transitions and resilience in ecological systems.
The theory of nucleation and growth describes how clusters are generated and spread out \cite{Rikvold_PRE1994}. Our study reveals the crucial effects of noise strength and system size on transition features. Generally, the stronger noise triggers the transition faster, and the larger system takes less time to recover compared with the smaller system under the same intensity of noise. For large systems or under strong noise, there are multiple clusters nucleated simultaneously in the beginning, and these clusters spread out to their neighbors and to the whole system. Our numerical simulations reveal that the transition time is narrowly centered at an average value, signaling a deterministic feature. For small systems or under weak noise, one cluster is generated first, which expands until the entire system finishes transitions. The transition times vary stochastically for different realizations of noise, and they universally follow an exponential distribution. 

\section{\label{sec:mathematical framework}Mathematical framework}

The current analytical framework for noise-induced transitions is specifically targeted at the low-dimensional system consisting of only a few interacting components (or only one variable) \cite{guttalImpactNoiseBistable2007, zengNoiseDelayinducedRegime2013, yangDelayNoiseInduced2014, forgostonPrimerNoiseInducedTransitions2018, alexandrovNoiseinducedTransitionsShifts2018}. It would be highly challenging to directly analyze the system consisting of many interacting components (for example, the square lattice). Before the investigation of multi-variable systems, it is helpful to understand the role of noise in single-variable systems and the possibility of transitions between alternative stable states.

\subsection{\label{subsec:single-variable system}Single-variable system}

\begin{figure}[H]
    \centering
    \includegraphics[width=0.6\linewidth]{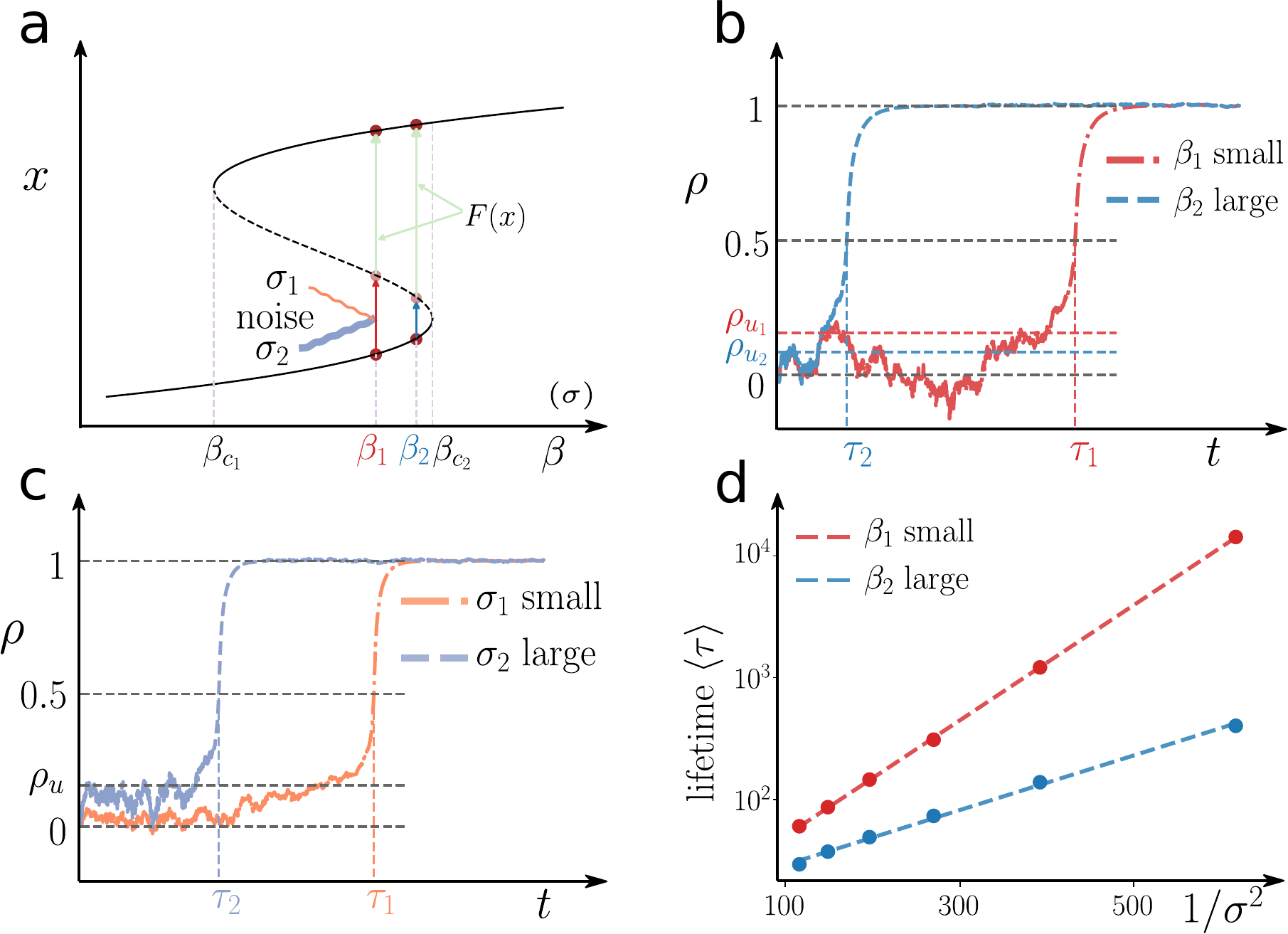}
    \caption{\textbf{The transition in the single-variable system.} (a) The resilience function of a general bistable system. For the bifurcation parameter $\beta \in (\beta_{c_1}, \beta_{c_2})$, there are two stable states ($x_\mathrm{L}$, $x_\mathrm{H}$) and one unstable state ($x_\mathrm{u}$). The initial stable state ($x_\mathrm{L}$) evolves to the unstable state ($x_\mathrm{u}$) by the aid of noise and is then naturally attracted to the other stable state ($x_\mathrm{H}$) by its deterministic dynamics. (b) and (c) display the evolution of the rescaled state $\rho$ in the presence of noise. (b) As $\beta_2$ is closer to the critical value $\beta_{c_2}$ than $\beta_1$, the unstable state $\rho_{\mathrm{u}_2}$ is lower, making the barrier in the landscape easier to cross in the presence of the same strength of fluctuations. The lifetime $\tau_2$ is thus smaller than $\tau_1$. (c) The parameter $\beta$ is the same, leading to the same landscape. In the presence of stronger noise $\sigma_2$, it is easier to drive the system to get over the barrier, causing $\tau_2$ to be smaller than $\tau_1$. (d) shows the simulation results of the average lifetime $\langle \tau \rangle $ under different values of $\beta$ and different noise strength. The dashed line is fitted based on Eq.~\myref{eq:tau_1D}. }
    \label{fig:F1}
\end{figure}

 For a single-dimensional system, the dynamics may follow the general form as
\begin{equation}
	\dv[]{x}{t}= f(x,\beta) + \eta(t).
	\label{eq:sde}	
\end{equation}
In Eq.~\myref{eq:sde}, $f(x, \beta) $ governs the deterministic dynamics as the tunable parameter $\beta $ captures the changing conditions, and $\eta(t)$ is delta-correlated noise with zero mean and variance $\langle \eta(t) \eta(t') \rangle  = \sigma ^2 \delta(t-t')$, where $\sigma$ is the standard deviation of the noise, which is referred to as noise strength. If the system has a fixed point, $x_0$, it satisfies $f(x_0, \beta) = 0$. This fixed point is stable if $\pdv[]{f(x, \beta)}{x}\Big\vert_{x_0}  < 0$, defined by the linear stability condition. These two equations enable us to derive the resilience function $x(\beta)$, which represents all the possible steady states 
of the system as a function of $\beta$, as shown in Fig.~\ref{fig:F1}a. This resilience function demonstrates that when the control parameter, $\beta$, is greater than $\beta_{c_2}$, the system only has one stable fixed point, denoted as $x_\mathrm{H}$, indicating a resilient state. The system will always recover to the fixed point, $x_\mathrm{H}$, for any state perturbations. When the parameter is less than $\beta_{c_1}$, the system also has a single fixed point, $x_\mathrm{L}$, indicating a collapse of the system. The system is not restorable, unless we change the parameter $\beta$. In this work, we particularly focus on the parameter space where $\beta$ is between $\beta_{c_1} $ and $\beta_{c_2}$ and close to one of the bifurcation points, such that the system has one unstable fixed point (the dashed line) and two stable fixed points, each with very different sizes of basin of attraction.
In the absence of noise,  to shift from $x_{\mathrm{L}}$ to $x_{\mathrm{H}}$, the environmental conditions would need to be restored all the way to $\beta_{c_2}$, to allow for a transition back to $x_{\mathrm{H}}$. However, noise of the appropriate intensity can trigger the transition without restoring the condition to $\beta_{c_2}$. For example, for a system with $\beta_1$ and the system is in the low state, this system can recover to the high state if the noise has a chance to push the system across the unstable fixed point. Furthermore, for two systems (the system 1 and the system 2) with parameters, $\beta_1$ and $\beta_2$ ($\beta_2 > \beta_1$), respectively, it is easier to restore the system 2, because the undesired state of the system 2 is closer to the unstable fixed point. Here, we introduce a normalized variable $\rho(t)$ in Eq.~\myref{eq:rho_single} to describe the state behavior during the restoration process. Thus, the normalized stable fixed points are $\rho_\mathrm{L}=0$, $\rho_\mathrm{H} = 1$ and the normalized unstable fixed point is $\rho_\mathrm{u}$ ($\rho_\mathrm{u}\in [0,1]$).

\begin{equation}
	\rho(t) = \frac{ x(t)  - x_\mathrm{L}}{ x_\mathrm{H} - x_\mathrm{L}}\label{eq:rho_single}
\end{equation}

In the following analysis, we focus on the case when the desired state $\rho_{\mathrm{H}}$ has a much larger basin of attraction than the undesired state $\rho_{\mathrm{L}}$, suitable to examine the ``one-way recovery'' from $\rho_{\mathrm{L}}$ to $\rho_{\mathrm{H}}$. Because of symmetry, the conclusions drawn from this manuscript should be applicable to the transition from $\rho_{\mathrm{H}}$ to $\rho_{\mathrm{L}}$ when $\rho_{\mathrm{L}}$ has a larger basin of attraction. (see Supplementary Information, Sec I for the detailed comparisons of one-way transitions and stochastic switching between stable states). For the case that we consider throughout the manuscript, the basin of attraction of the state $\rho_{\mathrm{L}}$ is much smaller than that of $\rho_{\mathrm{H}}$. In this case, the value of the unstable state $\rho_u$ is very close to $\rho_{\mathrm{L}}$, increasing the chance for perturbations to push the system from the state $\rho_{\mathrm{L}}$ over the unstable fixed point $\rho_u$ until the system gets attracted to the state $\rho_{\mathrm{H}}$. Notably, the backward transition is highly improbable by the same level of perturbations because of the much stronger attraction of the state $\rho_{\mathrm{H}}$. To quantify the time needed for the transition from $\rho_{\mathrm{L}}$ to $\rho_{\mathrm{H}}$, the half lifetime, $\tau$ (mathematically, $\tau = \{t| \rho(t) = 0.5\}$), can be safely used as an indicator of the degree of inertia associated with the transition from the undesired state to the desired one. One can choose any value as the cutoff as long as it is sufficiently larger than the unstable state $\rho_u$ \cite{kornissSpatialDynamicsInvasion2005, richardsMagnetizationSwitchingNanoscale1995}. 
The lifetime $\tau$ is determined by the noise strength $\sigma$ and the relative stability of alternative stable states controlled by $\beta$. Intuitively, larger noise strength indicates stronger fluctuations, which increases the chances of transition, making $\tau$ smaller (Fig.~\ref{fig:F1}c).
On the other hand, as $\beta$ increases, the basin of attraction of the stable state $\rho_{\mathrm{H}}$ gets larger, making the transition from $\rho_\mathrm{L}$ to $\rho_\mathrm{H}$ easier so that $\tau$ typically decreases (Fig.~\ref{fig:F1}b).
To better illustrate the transition process, the underlying landscape picture \cite{waddingtonStrategyGenesDiscussion1957} is introduced, and the effective potential energy is provided by Eq.~\myref{eq:V_eff} with zero potential energy at $x = 0$.
In the landscape representation, stable states are traditionally treated as valleys, and unstable states are pictured as hills \cite{schefferCatastrophicShiftsEcosystems2001, wellsControlStochasticInduced2015, xuexploringmechanismsdifferentiation2014}. The transition between stable states can be viewed as the transition from one shallow valley to a deeper valley by crossing the barrier.
	\begin{equation}
		V_\mathrm{{eff}}(x) = - \int _{0}^{x} \left[f(x')  \right]dx' \label{eq:V_eff}
	\end{equation}

	The transition time from one stable state to the other is a random variable because of the stochastic fluctuations, but the average value, $\langle \tau \rangle$, presents interesting properties. Following the derivations of the Kramers formula for the escape rate over a potential barrier by particles of Brownian motion \cite{kramersBrownianMotionField1940, tsimringNoiseInducedDynamicsBistable2001}, the average transition time $\langle \tau \rangle$ can be calculated through the analysis of the mean first passage time (MFPT), which is given by Eq.~\myref{eq:tau_1D}. From the quantitative predictions, one can notice that $\langle \tau \rangle$ increases exponentially with the potential energy difference $\Delta V= V(x_u) - V(x_\mathrm{L})$ between the low stable state and the unstable state (also interpreted as the barrier height) and decreases with noise intensity $\sigma ^2$, which is numerically verified in Fig.~\ref{fig:F1}d. The analysis of single-variable systems provides theoretical support for our intuitive assumption that the low barrier height and strong noise facilitate transitions, leading to resilience restoration see \cite{forgostonPrimerNoiseInducedTransitions2018, kramersBrownianMotionField1940} for derivation).
	\begin{equation}
		\langle \tau \rangle = \frac{2\pi}{ \sqrt{V''(x_\mathrm{L}) |V''(x_u)| } }  e^{2[V(x_u) - V(x_\mathrm{L}) ]/\sigma^2 } \label{eq:tau_1D}
	\end{equation}

	With the knowledge of the average lifetime $\langle \tau \rangle$ for single-variable systems available, we are ready to investigate transitions in multi-variable (spatially-extended) systems. Nucleation theory is utilized to analyze the generation and spread of the transition in multi-variable systems under external fluctuations.

\subsection{\label{subsec:spatially-extended system}Spatially-extended system}

Generally, the evolution of a system that consists of $N$ coupled components under external perturbations is described by Eq.~\myref{eq:multi-variable}. This study focuses on the spatially-extended ecosystem, where one type of species is considered, and its density varies in two-dimensional space and is discretized according to the square lattice topology with periodic boundaries. The deterministic dynamics of each node follows the same self-dynamics $F(x_i)$ and the same interaction dynamics $G(x_i, x_j )$, and parameters of these functions are also set uniformly for all components. The element of the adjacency matrix $A$ is either $0$ or a positive value $R$, which decides the coupling strength between interacting elements. 
Additionally, to model the external fluctuations acting on the node $i$, delta-correlated Gaussian noise $\eta_i(t)$ with zero mean and variance $\langle \eta_i(t) \eta_j(t') \rangle = \sigma ^2 \delta_{ij} \delta(t-t') $ is introduced, which is the same as the noise applied to the single-variable systems. 
This general framework can be used to describe a wide range of coupling systems under external perturbations.
\begin{equation}
	\dv[]{x_i}{ t} = F(x_i) + \sum _{ j =1}^{N  } A_{ij} G(x_i, x_j) + \eta_i(t) \label{eq:multi-variable}
\end{equation}

\begin{equation}
	\dv[]{x_{\mathrm{eff}}}{ t} = F(x_{\mathrm{eff}}) +  \beta_{\mathrm{eff}} G(x_{\mathrm{eff}}, x_{\mathrm{eff}}) \label{eq:reduction}
\end{equation}

Utilizing the dimensional reduction theory \cite{gaoUniversalResiliencePatterns2016}, the deterministic evolution of multi-variable systems can be reduced to Eq.~\myref{eq:reduction}, containing only one variable, $x_{\mathrm{eff}}$. 
In the lattice, the fixed states and their stability for each component are proved to be the same as in the reduced one-dimensional system \cite{jiangTrueNonlinearDynamics2020}.
Hence, according to the analysis of single-variable systems, for the specific values of effective interaction strength ($\beta \in (\beta_{c_1}, \beta_{c_2})$), each component in this system has two stable states and one unstable state. In the presence of noise, the transition from the low state $x_\mathrm{L}$ to the high state $x_\mathrm{H}$ is possible for every component. Once the first transition occurs to the node $i$, its neighbors will also recover through the interaction with it. 
To show the overall evolution properties of the entire system, the global state $\rho(t) $ is defined in Eq.~\myref{eq:global rho} by taking the average of the individual state value
\begin{equation}
	\rho(t)  = \langle  \rho_i(t)\rangle_N = \frac{1}{N} \sum _{i=1}^{N } \rho_i(t). \label{eq:global rho}
\end{equation}

The spread of such transition can be well described by the theory of homogeneous nucleation and growth in {\em finite} systems \cite{Rikvold_PRE1994, kornissSpatialDynamicsInvasion2005, omalleyInvasiveAdvanceAdvantageous2006}, and this theory can also predict the spatial-clustering pattern formed during the spreading process.
The transition from $\rho_{\mathrm{L}}$ to $\rho_{\mathrm{H}}$ occurs to some node at first, nucleating a cluster, and this cluster continues to expand until it fills the entire space, or the cluster’s edge reaches the edge(s) of other clusters that have nucleated in the system. Homogeneous nucleation makes two assumptions \cite{kornissSpatialDynamicsInvasion2005}: nucleation occurs in a Poisson process with a constant rate $I$ both temporally and spatially; once a cluster nucleates, it grows homogeneously with a {\em constant} radial velocity $v$. Since the interaction environments for all components are identical, and the perturbations they receive come from the same distribution, each node has the same chance to nucleate a cluster, which satisfies the assumptions of homogeneous nucleation. 

As predicted by homogeneous nucleation theory, the restoration process exhibits different patterns for small systems and large systems when the nucleation rate $I$ is fixed.
Small systems exhibit the single-cluster pattern because the number of candidates is so limited that the first cluster nucleates and spreads out to the rest of the system before the second possible cluster emerges. Since the nucleation for a specific individual follows a Poisson process, the global state $\rho$ is expected to evolve distinctly for different noise realizations. Also, the time to nucleate the first cluster, $t_n$, and the lifetime, $\tau$, are inherently random. We introduce the waiting time to quantify the time to the system recovery. Because the underlying process is modeled as the random Poisson process, the complementary cumulative probability distribution of waiting time $P_{\mathrm{not}}$ is derived as an exponential function, which represents the probability that the global state $\rho$ has not exceeded $0.5$ by time $t$. (Note that our chosen conventional cut-off value $0.5$ does not affect the findings.) 
The distribution of waiting time $P_\mathrm{not}$ is expressed as
	\begin{equation}
		P_{\mathrm{not}}(t) =
		\begin{cases}
			1, & t \le t_g\\
			e^{-(t -t_g) /\langle t_n \rangle }, & t>t_g 
		\end{cases}\;, \label{eq:P_not}
	\end{equation}
where $\langle  t_n \rangle $ is the average time elapsing until the first cluster nucleates (i.e., the first transition occurs); $t_g$ represents the time required for the global state $\rho $ to reach $0.5$ after the first cluster emerges, and this time depends on the linear size of the system, $t_g \sim N^{1/2}/ v$, where $v$ is the constant radial velocity. Also, $\langle t_n \rangle\sim (IN)^{-1}$, where $I$ is the nucleation rate per unit area. The average transition time from the initial undesired state to the desired state or the average lifetime of the initial state is expressed as $\langle \tau \rangle = \langle t_n \rangle  + t_g $. For small system sizes or in the weak-noise limit, the dominant term in the lifetime is the nucleation time, hence 
$\langle \tau \rangle\sim(IN)^{-1}$ (see Supplementary Information, Sec III for details).

In contrast, for large systems, more than one independent cluster nucleates and expands separately, leading to the multi-cluster mode. Spatial self-averaging reduces randomness of the global state $\rho$, making each individual $\tau$ closer to the average value and pushing $P_\mathrm{not}$ closer to a step function. In the large system-size limit, the lifetime distribution is narrowly centered about the average. Based on Avrami's Law, for sufficiently large systems \cite{Kolmogorov_1937,Johnson_1939, avramiKineticsPhaseChange1939, avramiKineticsPhaseChange1940,Ishibashi_JPSJ1971,Duiker_PRB1990}, the evolution of $\rho$ can be expressed in a deterministic form as $\rho = 1 - e^{ - \frac{\pi v^2 I }{3} t^3}$.
By setting $\rho=0.5$ (without loss of generality), the average lifetime for the multi-cluster mode can be obtained as 
$\langle \tau  \rangle  =   (\frac{3 \ln 2}{ \pi v^2 I } ) ^{1/3}$.
Then the evolution of $\rho$ can be described by
\begin{equation}
	\rho = 1 - e^{ - (\frac{t}{ \langle \tau \rangle  }) ^3\ln 2}.\label{eq:rho_large}
\end{equation}

According to Avrami's Law and homogeneous nucleation in finite systems \cite{Rikvold_PRE1994,kornissSpatialDynamicsInvasion2005, omalleyInvasiveAdvanceAdvantageous2006}
, the average lifetime of two transition modes is summarized as
	\begin{equation}
		\langle \tau \rangle  \sim
		\begin{cases}
			 \frac{1}{IN}, &  N^{\frac{1}{2}} \ll R_0 \quad\textrm{(single-cluster mode)}\\
			 I^{-\frac{1}{3}}, & N^{\frac{1}{2}} \gg R_0  \quad\textrm{(multi-cluster mode)}
		 \end{cases}\;, \label{eq:tau_two_modes}
	\end{equation}
where $R_0 \sim  (\frac{v}{I})^{1/3} \sim I^{-1/3}$ is the typical distance between separate clusters (and $N^{1/2}$ is the linear size of the two-dimensional lattice).
Transition patterns for different system sizes and nucleation rates are classified into two distinct cluster nucleation modes, separated by a {\em crossover} region centered around the curve, $N^{1/2} \sim R_0$.
Small systems or low nucleation rates induce the single-cluster mode, while large systems or high nucleation rates exhibit the multi-cluster mode.

By constructing a scaling function \cite{OMalley_PhD2008} with the following asymptotic behavior,
\begin{equation}
	S(u) \sim
	\begin{cases}
		u^2, & u \gg 1  \quad\textrm{(single-cluster mode)}    \\
        \mathrm{const.}, &u\ll 1  \quad\textrm{(multi-cluster mode)}
	\end{cases} \;,\label{eq:scaling}
\end{equation}
where $u=R_0/N^{1/2}\sim v^{1/3}/(I^{1/3}N^{1/2}) \sim I^{-1/3}N^{-1/2}$, one can capture the average lifetime of {\em any} system size and nucleation rate (including the {\em crossover} between the single-cluster and multi-cluster regimes),
	\begin{equation}
		\langle \tau  \rangle = I^{-\frac{1}{3}} S(R_0/N^{\frac{1}{2}}) =  I^{-\frac{1}{3}} S(I^{-\frac{1}{3}} N^{-\frac{1}{2}})\label{eq:Scaling_Avrami} \;.
	\end{equation}

Motivated by the study on the average transition time $\langle \tau \rangle $ for single-variable systems [Eq. (\ref{eq:tau_1D})], the relationship between local nucleation rate and noise strength in spatially-extended systems is expected to scale as
\begin{equation}
	I  \sim e^{-\frac{c}{\sigma^2 }},\label{eq:nucleation rate}
\end{equation}
where $c$ is a constant specific to the given dynamics and can be empirically fitted in the weak-noise case.
In turn, the average lifetime $\langle \tau \rangle$ for different system sizes and noise strength can be described by a universal scaling function: employing Eqs.(\ref{eq:Scaling_Avrami}) and (\ref{eq:nucleation rate}), when plotting
$\langle \tau  \rangle e^{-\frac{c}{3\sigma^2}}$ vs. $e^{\frac{c}{3\sigma^2 }}/N^{\frac{1}{2}}$,
all curves should collapse on the same curve $S(u)$.

\section{\label{subsec:mutualistic system} Resilience restoration of mutualistic systems}

To verify the predictions by nucleation theory of the noise-induced transition patterns in systems with alternative stable states, we first study resilience restoration of the mutualistic system. 

\subsection{Dynamical model}
We use Eq.~\myref{eq:mutualistic} as the deterministic dynamics to track the abundance of one species distributed among a square lattice in the mutualistic system \cite{hollandPopulationDynamicsMutualism2002, suweisEmergenceStructuralDynamical2013}. The self-dynamics $F(x_i)$ describes that the growth of the species in each location follows the logistic law with the Allee effect, and the dynamics $G(x_i, x_j)$ accounts for the mutualistic interaction between the species in two neighboring locations $i$ and $j$ through the interaction strength $A_{ij}$ defined in Eq.~\myref{eq:multi-variable}. 
\begin{equation}
\begin{split}
	&F(x_i) = B_i   + x_i  \left(  1- \frac{x_i}{ K_i} \right) \left(\frac{x_i}{ C_i} - 1 \right) \\
	&G(x_i, x_j ) = \frac{x_i x_j }{D_i + E_i x_i + H_j x_j} \label{eq:mutualistic}.
\end{split}
\end{equation}

Note that we use the same parameters as Ref. \cite{gaoUniversalResiliencePatterns2016}. The parameters are node-uniform and set as $B_i = B = 0.1 $, $C_i = C = 1$, $D_i = D = 5$, $E_i = E = 0.9$,  $H_j = H = 0.1$, $K_i = K = 5$, and the interaction strength $R=1$ (leading to the effective interaction strength $\beta_{\mathrm{eff}}=4$).
At some moment, if the species in all locations are trapped in the low stable state $x_{\mathrm{L}}$, they will stay at this state forever if there is no action or perturbation. 
It is, for sure, not desired from the ecological viewpoint.
If one would like to keep the system always in the high stable state $x_\mathrm{H}$, a straightforward approach is to increase the interaction strength to ensure that $\beta_{\mathrm{eff}}$ is larger than the critical bifurcation value $\beta_{c2} $. In this case, the system will be attracted to the high state no matter where it starts. 
Alternatively, noise has shown the ability to induce the transition between two stable states from the study of one-variable systems. We are particularly interested in how noise assists the resilience restoration (i.e., transferring the system from the undesirable state $x_\mathrm{L}$ to the desired state $x_\mathrm{H}$). 

\subsection{Patterns of resilience restoration}
\begin{figure}[H]
    \centering
    \includegraphics[width=1\linewidth]{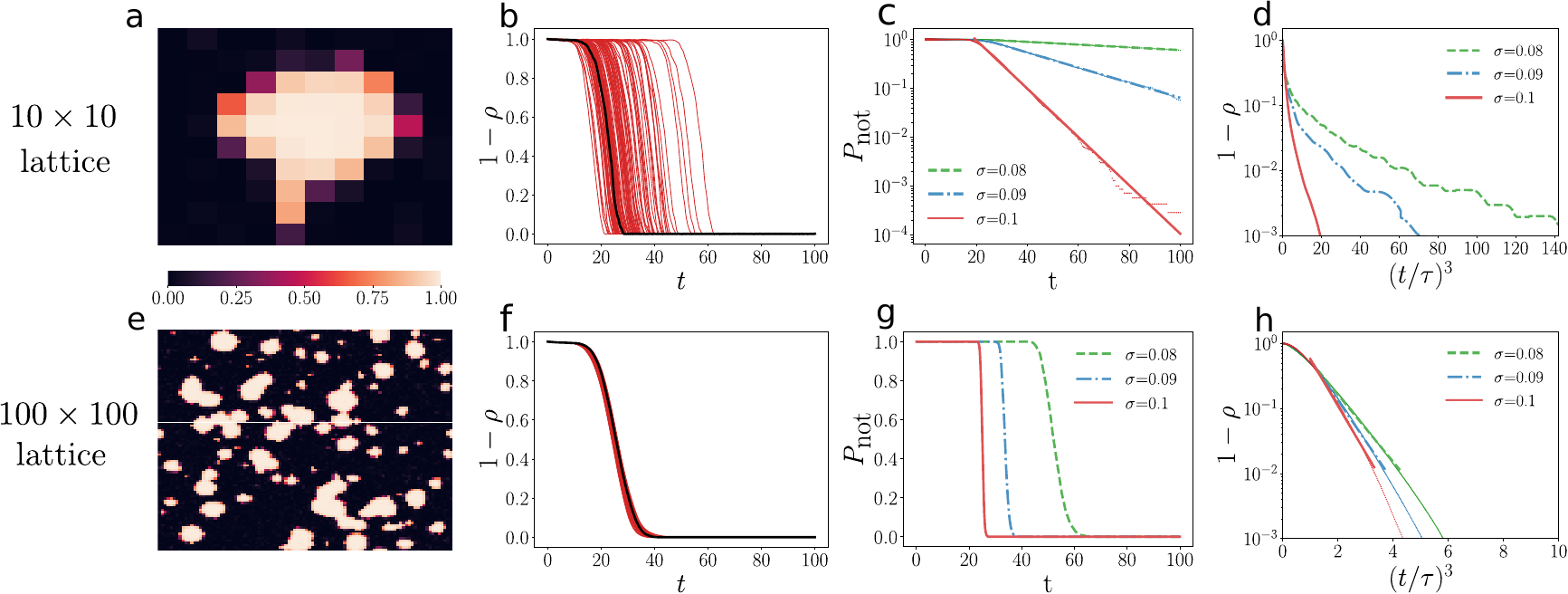}
    \caption{\textbf{Single-cluster and multi-cluster modes in the mutualistic system.}  
    (a) One snapshot for the state $\rho_i$ of each node under the noise of the standard deviation $\sigma = 0.1$. Initially, all of the nodes are at the low state $x_{\mathrm{L}}$. At some time later, the transition to the high state $x_{\mathrm{H}}$ occurs to one node, which is treated as a single cluster. Such transition then spreads out to its neighbors. (b) The evolution of the global state $\rho$ for $100$ realizations of the single-cluster mode (a). (c) The probability distribution of waiting time $P_{\mathrm{not}}$ for the fixed system size and the various noise strengths $\sigma=0.08, 0.09, 0.1$. The single dots are simulation data, and different types of lines are obtained by linear fit according to Eq.~\myref{eq:P_not}.  (d) The evolution of the average global state $\rho $ using the same data as in (c).
    (e) One snapshot shows the state $\rho_i$ of each node for the case when all nodes starts from $x_{\mathrm{L}}$ initially, and $\sigma$ is also $0.1$. Different from (a), the transition to $x_{\mathrm{H}}$ occurs at several separate nodes, and they expand independently, forming the multiple cluster. (f) The evolution of the global state $\rho$ for $100$ realizations, which are more centered around a certain value instead of being random in (b). (g) The distribution $P_\mathrm{not}$ for $N = 10000$ and $\sigma = 0.08, 0.09, 0.1$, which approaches the step function as $\sigma$ increases. (h) The evolution of the global state $\rho $ averaged over $100$ realizations using the same data as in (g). The single dots are simulation data, and different types of lines are obtained by linear fit according to Eq.~\myref{eq:rho_large}.}
    \label{fig:F2}
\end{figure}

Let us consider the case when all nodes in the system start from the low state $x_{\mathrm{L}}$, indicating a collapsed state.
Observed from simulations, the proper noise can excite some node to $x_{\mathrm{H}}$, and such transition spreads out to its neighbors via interaction until the rest of the system completes transitions.
According to homogeneous nucleation theory, for different system sizes and nucleation rates, two possible transition patterns are present.
We successfully show the snapshots of two cluster modes by numerical simulations, the single-cluster and multi-cluster mode.
Notably, these two modes possess radically distinct properties. For the single-cluster mode (Fig.~\ref{fig:F2}a), there is only one node that switches from $x_{\mathrm{L}}$ to $x_{\mathrm{H}}$ in the beginning, and the rest of nodes shift to $x_{\mathrm{H}}$ through the interactions with the neighbors which have already transitioned; while for the multi-cluster mode (Fig.~\ref{fig:F2}e), there is more than one node in the separate location that transfers to $x_{\mathrm{H}}$ simultaneously. It is expected since for the large system, there are enough candidates to receive fluctuations, increasing the chance to induce independent transitions.

As predicted by nucleation theory, the evolution of $\rho(t)$ for the single-cluster mode and multi-cluster mode differs a lot, and numerical results confirm the difference. Fig.~\ref{fig:F2}b and Fig.~\ref{fig:F2}f display $100$ realizations for system sizes $N =100$ and $N=10000$ under the same intensity of noise. For the single-cluster mode (Fig.~\ref{fig:F2}b), the evolution varies for individual realizations, so the transition times are different, indicating the uncertain evolution feature. In contrast, for the multi-cluster mode (Fig.~\ref{fig:F2}f), the evolution of $\rho(t)$ is similar for different realizations, implying that the evolution is deterministic in the infinite system-size limit.
Following this, the waiting time distribution $P_\mathrm{not}$ for two cluster modes is also verified (Fig.~\ref{fig:F2}c). For the single-cluster mode, $P_\mathrm{not}$ is initially constant and then decreases exponentially with respect to $t$ verifying Eq.~\myref{eq:P_not}. The slope of the distribution gets more negative as noise becomes stronger suggesting a larger nucleation rate.
For the multi-cluster mode, $P_\mathrm{not}$ gets closer to a step function (Fig.~\ref{fig:F2}g) as noise strength increases. This is because the larger nucleation rate induces more separate clusters for a given system size and thus leads to the more deterministic evolution by self-averaging.
From the theoretical analysis, the global state $\rho$ for the multi-cluster mode evolves predictably according to Eq.~\myref{eq:rho_large}. Whereas, for the finite-size system, the evolution of the multi-cluster mode (Fig.~\ref{fig:F2}h) is not perfectly deterministic, but still much less random than the single-cluster mode (Fig.~\ref{fig:F2}d). 

\subsection{The role of system size and noise strength}

\begin{figure}[H]
    \centering
    \includegraphics[width=0.6\linewidth]{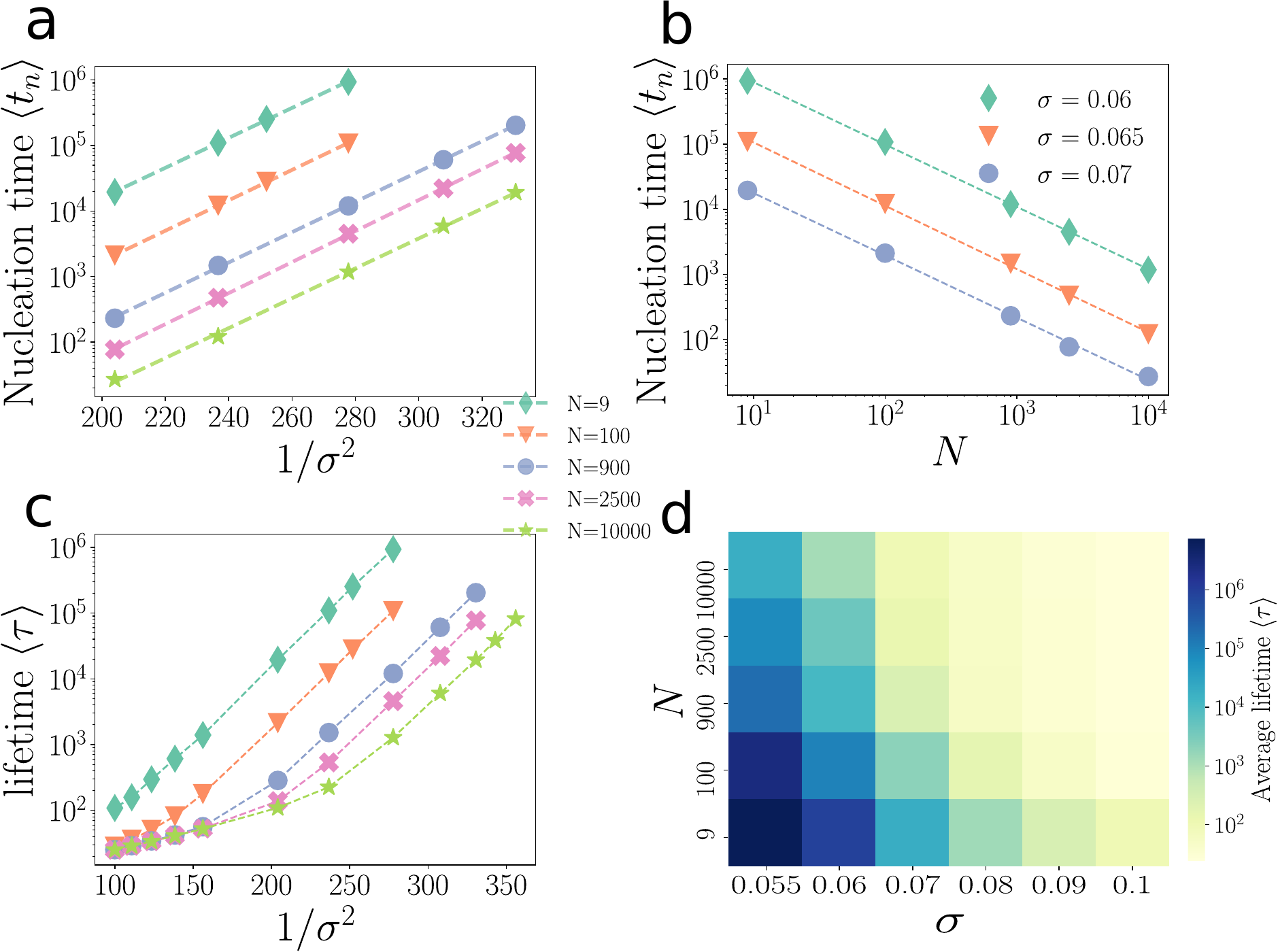}
    \caption{\textbf{The influence of system size $N$ and noise strength $\sigma$ on transition modes and average lifetime.} Initially, all of the nodes are at the low state $x_{\mathrm{L}}$, and the time to switch to the high state $x_{\mathrm{H}}$ is measured. (a) The average nucleation time $\langle t_n \rangle  $ changes with noise strength for different system sizes. (b) The linear relationship between $ \langle \tau \rangle $ and $N^{-1}$. (c) Two regimes with different slopes of  $\ln \langle \tau \rangle $ as a function of $\sigma^{-2}$ corresponding to two cluster modes. (c) and (d) summarize the effects of system size $N$ and noise strength $\sigma$ on the average lifetime $\langle \tau \rangle $. The increase of noise strength lowers the average lifetime. For the single cluster mode, the larger system requires less time to complete transitions.}
  \label{fig:F3}
\end{figure}

The cluster mode not only depends on the system size but also relies on the nucleation rate, which is decided by noise strength. 
Typically, large systems under strong noise belong to the multi-cluster mode, while small systems under weak noise exhibit the single-cluster mode. 
However, low nucleation rates resulting from weak noise can induce the single-cluster mode even for a very large system (Fig.~\ref{fig:F3}a, \ref{fig:F3}b).
Also, the average nucleation time $\langle  t_n \rangle = (NI)^{-1} $ for the single-cluster mode is validated.
The average lifetime $\langle \tau \rangle $ behaves differently for the two cluster modes and exhibits two regimes (Fig.~\ref{fig:F3}c), and its value increases exponentially as $\sigma^{-2}$ increases for both cluster modes as predicted by Eq.~\myref{eq:tau_two_modes}. For the given dynamics, $\langle \tau \rangle$ entirely depends on the system size $N$ and noise strength $\sigma$ (Fig.~\ref{fig:F3}c, \ref{fig:F3}d), and there is a decrease as $N$ or $\sigma$ increases. One can also notice that the slope of $\ln \langle \tau \rangle $  as a function of  $\sigma^{-2}$ for the single-cluster mode is larger than the slope for the multi-cluster mode.

\subsection{Phase diagram and finite-size scaling}
\begin{figure}[H]
    \centering
    \includegraphics[width=0.6\linewidth]{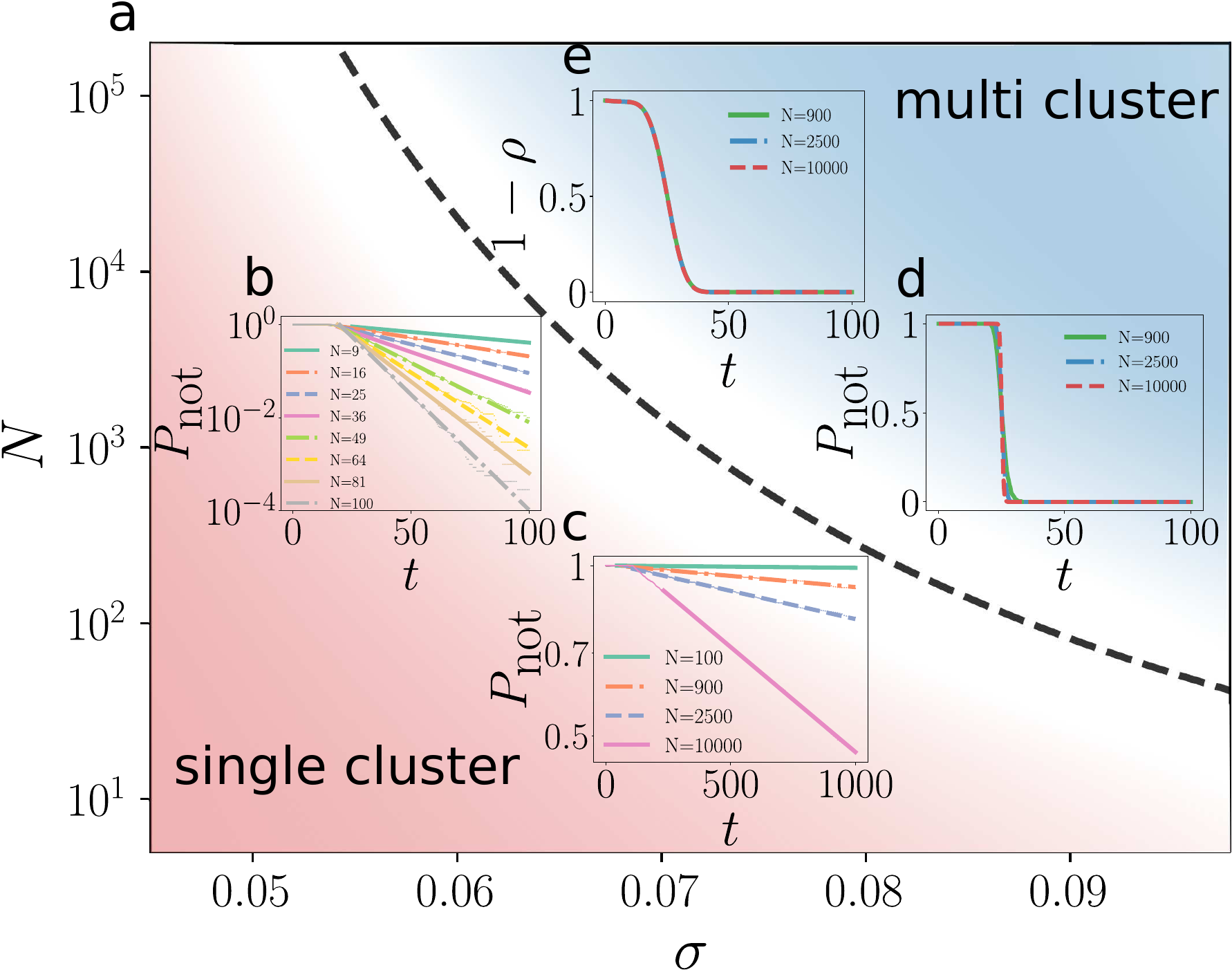}
    \caption{\textbf{Crossover between two cluster modes for the sample mutualistic system.} Initially, all of the nodes are at the low stable state $x_{\mathrm{L}}$, and they are driven to the high stable state $x_{\mathrm{H}}$ in the presence of noise. (a) According to Eq.~\myref{eq:tau_two_modes}, two cluster modes are distinguished. The dashed curve is drawn according to the equation $N^{1/2} = e^{\frac{c}{3\sigma^2}}$, where $c$ is a fitted parameter. The gradual change of the background color (red-white-blue) is to qualitatively illustrate the continuous nature of the crossover from the single-cluster mode to the multi-cluster mode. The dashed curve corresponds to the center of the crossover region (provided by the above formula), separating the two cluster-growth modes. (b) and (c) describe the single-cluster mode (in red), while (d) and (e) display the multi-cluster mode (in blue). (b) The distribution $P_\mathrm{not}$ for the fixed noise strength $\sigma=0.1$ and different system sizes $N=9, 16, 25, 36, 49, 64, 81, 100$. (c) $P_\mathrm{not}$ for the fixed weak noise $\sigma = 0.06$ and $N = 100, 900, 2500, 10000$. In both (b) and (c), the single dots are simulation data, and different types of lines are obtained by linear fit according to Eq.~\myref{eq:P_not}. (d) $P_\mathrm{not}$ for the fixed noise strength $\sigma=0.1$ and different system sizes $N=900, 2500, 10000$. (e) The evolution of the global state $\rho$ using the same data as in (d).}
  \label{fig:F4}
\end{figure}

In accord with the scaling theory, Eqs. (\ref{eq:tau_two_modes})-(\ref{eq:Scaling_Avrami}), the two distinct cluster-growth modes are separated by a {\em crossover} region centered around $N^{1/2} \sim R_0$ (Fig.~\ref{fig:F4}a). One should note that this crossover (centered around the dashed curve in Fig.~\ref{fig:F4}a) is not a sharp transition separating the two cluster-growth modes. The gradual change of the background color (red-white-blue) is to qualitatively illustrate the continuous nature of the crossover from the single-cluster mode to the multi-cluster mode.
The small system (Fig.~\ref{fig:F4}b) or weak noise induces (Fig.~\ref{fig:F4}c) the single-cluster mode, while the large system or relatively strong noise (Fig.~\ref{fig:F4}d, \ref{fig:F4}e) produces the multi-cluster mode.
According to the proposed scaling function Eq.~\myref{eq:Scaling_Avrami}, employing and plotting properly scaled variables, $\langle \tau  \rangle e^{-\frac{c}{3\sigma^2}}$ vs. $e^{\frac{c}{3\sigma^2 }}/N^{\frac{1}{2}}$, we expect that all numerical data would collapse onto the scaling function $S(u)$, capturing general nucleation behavior \cite{OMalley_PhD2008}.

\subsection{The universal scaling law}
\begin{figure}[H]
    \centering
    \includegraphics[width=1\linewidth]{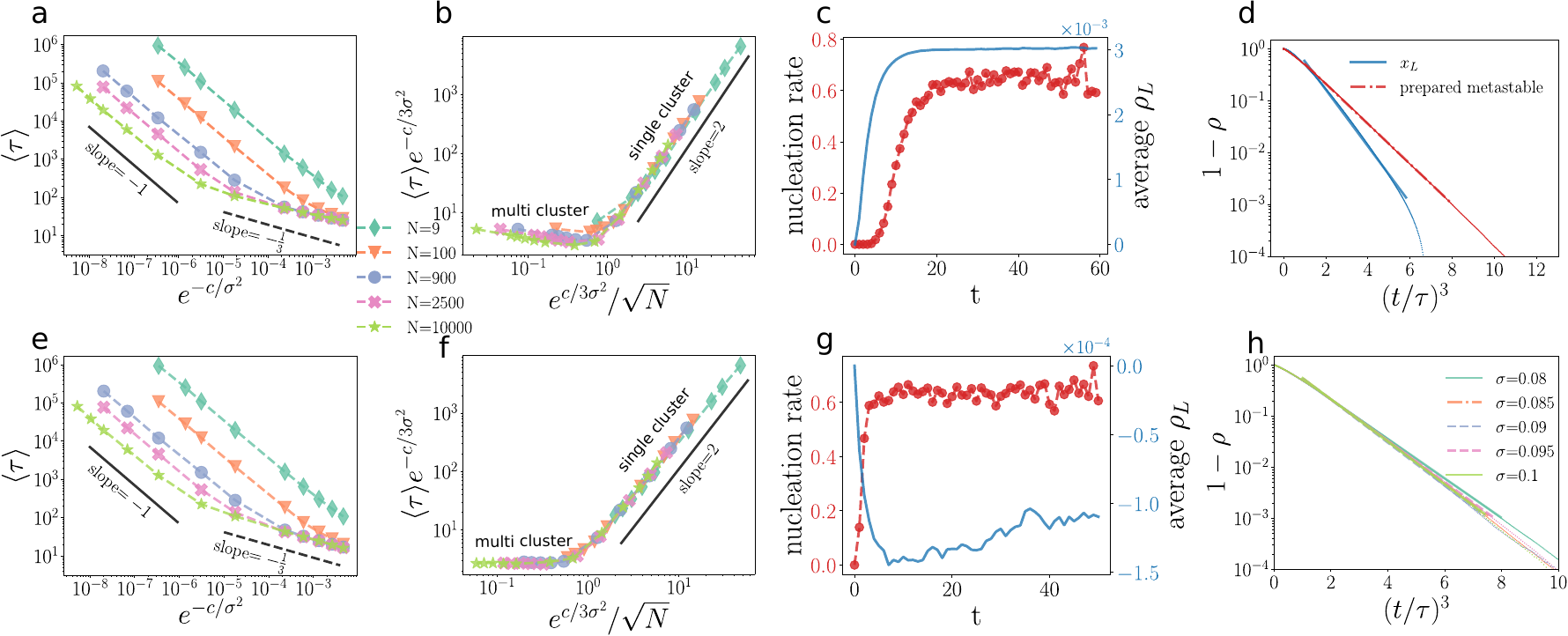}
    \caption{\textbf{Scaling between two cluster modes in the mutualistic system.} For (a) -- (c), the system starts from the low state $x_{\mathrm{L}}$, and the time to reach the state $x_{\mathrm{H}}$ is characterized by $\tau$. (a) The relationship of $\langle \tau \rangle $ and $e^{-\frac{c}{\sigma^2}}$ differs between two cluster modes. (b) The finite-size scaling is drawn by assuming the slope of multi-cluster mode in (a) is $-\frac{1}{3}$. For (c) and (d), $N=10000$ and $\sigma=0.08$. (c) The nucleation rate increases before the average state of nodes which have not transitioned stabilizes. (d) The system starts to evolve from $x_\mathrm{L}$ and the prepared state, respectively. The evolution of the global state $\rho$ for the latter case agrees better with Eq.~\myref{eq:rho_large} than the former case. For (e) -- (h), the system starts from the prepared metastable state, and the time to reach the state $x_{\mathrm{H}}$ is recorded as $\tau$. (e) The average lifetime $\langle \tau \rangle $ for two cluster modes. (f) The finite-size scaling is consistent with the theoretical prediction in Eq.~\myref{eq:Scaling_Avrami}. (g) The nucleation rate needs less time to stabilize compared with (c). (h) The evolution of the global state $\rho$ for the multi-cluster mode when $N =10000$ and $\sigma = 0.08, 0.085, 0.09, 0.095, 0.1$. In both (d) and (h), the single dots are simulation data, and different types of lines are obtained by linear fit according to Eq.~\myref{eq:rho_large}.} 
    \label{fig:F5}
\end{figure}

Nevertheless, observed in Fig.~\ref{fig:F5}a, for the multi-cluster mode, $\langle \tau \rangle $ is not precisely proportional to $I^{-1/3}$. Therefore, the two transition modes cannot be scaled in a satisfactory fashion as the scaled data for the multi-cluster regime is not constant (Fig.~\ref{fig:F5}b), which contradicts the assumption of Eq.~\myref{eq:scaling}.
The deviation from Avrami's Law suggests that the assumption(s) of homogeneous nucleation might be violated. For a large system subjected to relatively strong noise, which guarantees the multi-cluster mode, the nucleation rate changes in the beginning (Fig.~\ref{fig:F5}c), defying the assumption of constant nucleation rate. The initial rise of the spatially distributed low state for the nodes which have not been driven to the high state indicates that the initial state is not a metastable state. In the presence of noise, the metastable state is a state distribution where all nodes are around the low state $\rho_{\mathrm{L}}$, but some nodes are closer to the unstable state $\rho_u$ and some are even below the state $\rho_{\mathrm{L}}$. The system in the presence of noise needs time to reach a new metastable state which is different from the low stable state. To keep the nucleation rate constant, one should use the metastable configuration as the initial condition. However, as seen in Fig.~\ref{fig:F5}c, some cluster have already emerged before the system reaches the metastable configuration indicated by the increase of the average low state and the nucleation rate, so that the preprocessing of the initial state is needed to forbid any transition before the metastable configuration is constructed. One can artificially prepare the system close to the metastable state by reverting the node back to $\rho_\mathrm{L}$ if any other sites nucleate during preprocessing.

To gain further insight into the source of this discrepancy, we carried out simulations with the initial configurations being very close to the metastable state. In such a case, the nucleation rate $I$ stabilizes much faster (Fig.~\ref{fig:F5}g) compared with the system without preprocessing. Afterward, the nucleation rate is roughly constant, so homogeneous nucleation theory can be reliably applied. This leads to a better agreement between the simulation and the theory, as the evolution of $\rho$ is closer to what Eq.~\myref{eq:rho_large} postulates (Fig.~\ref{fig:F5}d, \ref{fig:F5}h), and the data from the two cluster modes follows the scaling function (Fig.~\ref{fig:F5}e, \ref{fig:F5}f). It is expected that the average lifetime $\langle \tau \rangle$ agrees better with Eq.~\myref{eq:tau_two_modes} if the metastable state can be perfectly prepared.

\section{\label{subsec:diffusion dynamics} Restoration of diffusion dynamics}

To further validate the proposed theory, we adapt three well-studied ecological models exhibiting alternative stable states \cite{dakosSpatialCorrelationLeading2010, chenEigenvaluesCovarianceMatrix2019} with diffusive interaction and then apply the above theory to investigate the transition features.

\subsection{Examples of some dynamics}
The self-dynamics for three diffusion models are defined below.

The harvesting model in Eq.~\myref{eq:harvesing} describes the growing resource biomass with fixed grazing rate \cite{mayThresholdsBreakpointsEcosystems1977}. The first term on the right hand side of Eq.~\myref{eq:harvesing} describes logistic growth, where $r$ is the maximum growth rate, and $K$ is the carrying capacity. The second term is the ``Holling's type \Romannum{3}" consumption function \cite{hollingCharacteristicsSimpleTypes1959}. The system transitions from an underexploited state to overexploited state as the harvesting rate $\beta$ exceeds a certain critical value.

\begin{equation}
	F(x_i) = r x_i (1- \frac{x_i}{K} ) - \beta \frac{x_i^2}{x_i^2 + 1} \label{eq:harvesing}
\end{equation}

The eutrophication model in Eq.~\myref{eq:eutrophication} describes the dynamics of nutrient concentration in the eutrophic lake \cite{carpenterManagementEutrophicationLakes1999}. The variable $x_i$ represents the density of phosphorus mass (nutrient) in the location $i$ of the lake. The first term $a$ on the right hand side of Eq.~\myref{eq:eutrophication} is the nutrient loading rate, the second term describes nutrient loss processes with the rate $r$, and the last term accounts for recycling processes following a sigmoid function.  As the maximum recycling rate $\beta$ increases to the critical point, the lake transfers from oligotrophic to eutrophic.

\begin{equation}
	F(x_i) = a - r x_i + \beta \frac{x_i^8}{x_i^8 + 1} \label{eq:eutrophication}
\end{equation}

The vegetation-turbidity model in Eq.~\myref{eq:vegetation} describes the vegetation dynamics considering turbidity \cite{schefferStoryShallowLakes2004}. The variable $x_i$ represents the density of aquatic vegetation in the location $i$ of the lake. The first term on the right hand side of Eq.~\myref{eq:vegetation} characterizes the growth of vegetation with the maximum growth rate $r_v$. The function $E$ is an inverse Monod function and used to describe the vegetation effect on turbidity. Accordingly, the second term is a Hill function describing the sigmoidal decline of vegetation
with turbidity. 
 As the water becomes turbid, indicated by the background turbidity $\beta$, macrophytes suddenly decrease.

\begin{equation}
\begin{split}
	F(x_i) &= r_v x_i  - r_v x_i^2 \frac{r^4 + E_i ^4}{r^4} \label{eq:vegetation}\\
 	E_i& = 	\frac{h_v \beta}{h_v + x_i}
\end{split}
\end{equation}

All three models can exhibit alternative stable states with the properly chosen bifurcation parameter $\beta$ as illustrated in Fig.~\ref{fig:F7}a, \ref{fig:F7}e, \ref{fig:F7}i. The transition from the low stable state to the high stable state is possible in the presence of noise for such single-variable systems.

\subsection{The spatially-extended diffusion dynamics}

Likewise, the underlying topology is also a square lattice with periodic boundaries. The interaction dynamics is defined in Eq.~\myref{eq:diffusion_G}, which represents the diffusive process between adjacent neighbors, and the interaction strength $R$ determines the uniform diffusion rate. The species density at each location varies according to the internal dynamics as defined in Eqs.~\myref{eq:harvesing} -- \myref{eq:vegetation} , and it is also influenced by dispersion to or immigration from neighbors and stochastic environmental fluctuations.

\begin{equation}
	G(x_i, x_j) = x_j - x_i, \label{eq:diffusion_G}
\end{equation}

Once the system gets stuck in a malfunctioning state, resilience restoration is required. 
In the presence of noise, each component is likely to be driven from the undesired state to the functional stable state, and then the entire system undergoes substantial changes due to the transition of one or a few nodes. Nucleation theory can be employed as well to study the overall transition features.
The results we collected from three diffusion models are similar to those obtained from the mutualistic system and thus verify the predictions by nucleation theory.
We consider the case when the undesired state has a much smaller basin of attraction than the desired state so that noise can drive the system from the initial undesired state to the functional state, resulting in resilience restoration.

\subsection{The phase diagram of Harvesting system}
\begin{figure}[H]
    \centering
    \includegraphics[width=0.7\linewidth]{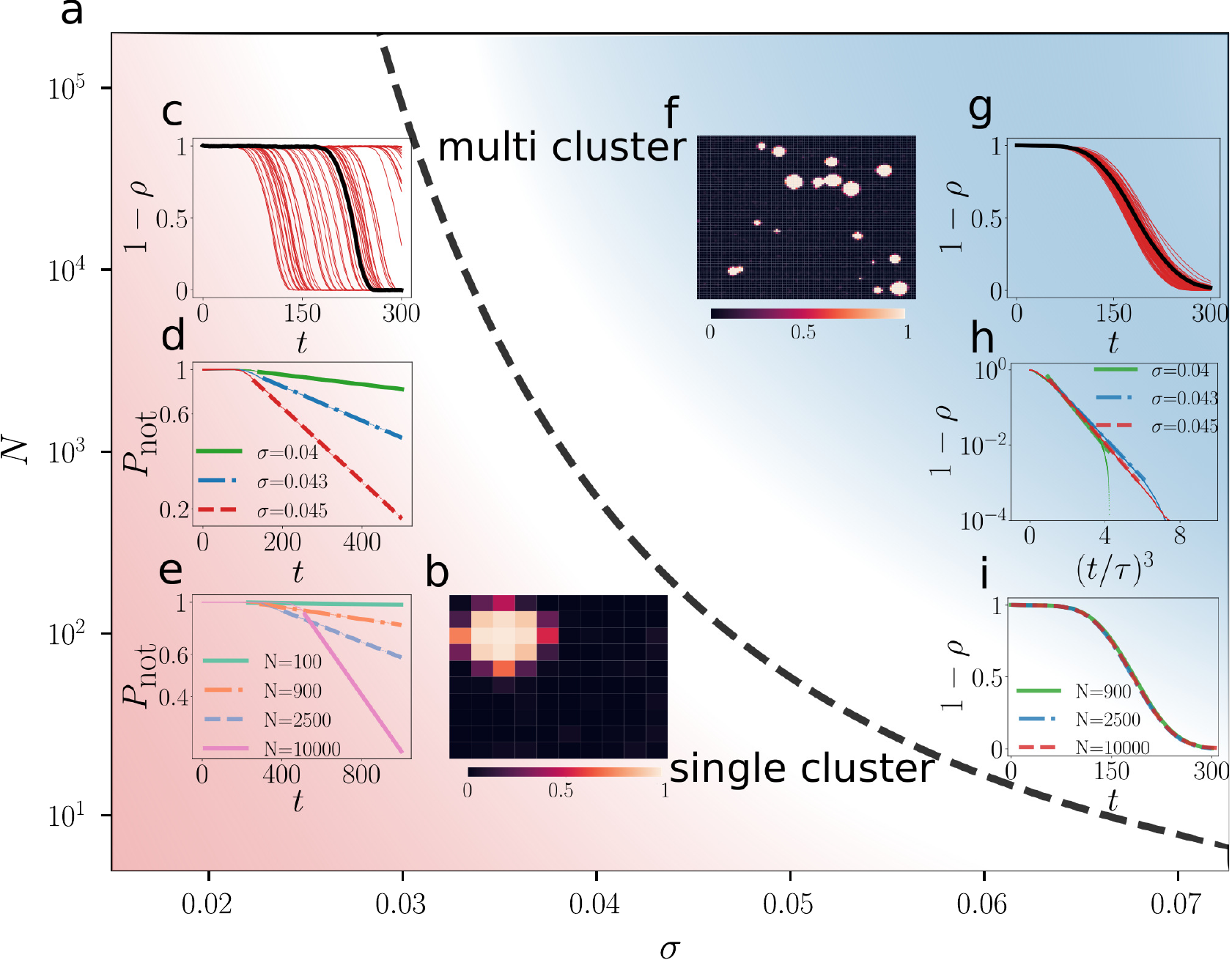}
    \caption{\textbf{Harvesting system.} The parameters are set as $r=1$, $K=10$, the diffusion rate $R= 0.02$, and the bifurcation parameter $\beta= 1.8$. Initially, all of the nodes are at the low stable state $x_{\mathrm{L}}$, and they are driven to the high stable state $x_{\mathrm{H}}$ in the presence of noise. (a) Two clusters are separated according to the curve $N^{1/2} = e^{\frac{c}{3\sigma^2}}$, where $c$ is a fitted parameter. As in Fig.~\ref{fig:F4}, the gradual change of the background color (red-white-blue) is to qualitatively illustrate the continuous nature of the crossover from the single-cluster mode to the multi-cluster mode. The dashed curve corresponds to the center of the crossover region (provided by the above formula), separating the two cluster-growth modes. For (b) -- (d), the system size $N = 100$ ($10 \times 10 $ lattice). (b) One snapshot shows the evolution of each node $\rho_i$ in the presence of noise of the standard deviation $\sigma = 0.045$. The initial states for all of the nodes are $x_{\mathrm{L}}$. At some time later, the transition to $x_{\mathrm{H}}$ occurs to one node, which is treated as a single cluster, and it spreads out to its neighbors. (c) The evolution of the global state $\rho$ for $100$ realizations, which corresponds to the single-cluster mode in (a). (d) The distribution of waiting time $P_{\mathrm{not}}$ for the fixed system size and the varied noise strengths $\sigma=0.04, 0.043, 0.045$. (e) $P_\mathrm{not}$ for the fixed weak noise $\sigma = 0.035$ and different system sizes $N = 100, 900, 2500, 10000$. In both (d) and (e), the single dots are simulation data, and different types of lines are obtained by linear fit according to Eq.~\myref{eq:P_not}. For (f) -- (h), the system size $N=10000$ ($100 \times 100$ lattice). (f) The snapshot illustrates the evolution of the system starting from $x_{\mathrm{L}}$, where $\sigma$ is also $0.045$. Different from (b), the transition to $x_{\mathrm{H}}$ occurs at several separate nodes, and they expand independently, forming the multiple cluster. (g) $100$ realizations of the global state $\rho$, which are more centered around a certain value instead of being random in (c). (h) The evolution of $\rho $ averaged over $100$ realizations for $\sigma = 0.04, 0.043, 0.045$. The single dots are simulation data, and different types of lines are obtained by linear fit according to Eq.~\myref{eq:rho_large}. (i) The evolution of the global state $\rho $ averaged over $100$ realizations for the fixed strong noise $\sigma=0.045$ and large system sizes $N=900, 2500, 10000$.}
    \label{fig:F6}
\end{figure}

Take the harvesting model as an example to illustrate the successful application of nucleation theory to resilience restoration.
Two cluster-growth modes are observed and separated by a crossover around $N^{1/2} \sim R_0$ (Fig.~\ref{fig:F6}a). Similar to the mutualistic system, the large system or strong noise produces multi-cluster mode; conversely, the small system or weak noise induces single-cluster mode.
For the small system ($N=100$ in the example), there is only one cluster formed during the transition (Fig.~\ref{fig:F6}b), exhibiting the single-cluster pattern. The individual lifetime $\tau$ in Fig.~\ref{fig:F6}c varies a lot for different realizations, indicating a stochastic feature. As derived above, the distribution of waiting times $P_\mathrm{not}$ follows an exponential function after a certain period $t_g$, observed in Fig.~\ref{fig:F6}d, \ref{fig:F6}e.
If the system is exposed to weak noise which means a low nucleation rate, it is very likely to enter the single-cluster regime even the size is sufficiently large (Fig.~\ref{fig:F6}e).
For the large system (like $N=10000$) with proper intensity of noise, more than one node in the separate location is recovered simultaneously (Fig.~\ref{fig:F6}f), presenting a multi-cluster pattern. Spatial-averaging reduces randomness so that the individual lifetime $\tau$ is more centered about a certain value (Fig.~\ref{fig:F6}g). The evolution is much more deterministic than the system in a single-cluster mode. The global state $\rho$ evolves approximately as Eq.~\myref{eq:rho_large} predicts (Fig.~\ref{fig:F6}h). If the applied noise is strong enough, the moderate-sized system  can still exhibit the multi-cluster mode (Fig.~\ref{fig:F6}i).

\subsection{The universal scaling in diffusion dynamics}

\begin{figure}[H]
    \centering
    \includegraphics[width=1\linewidth]{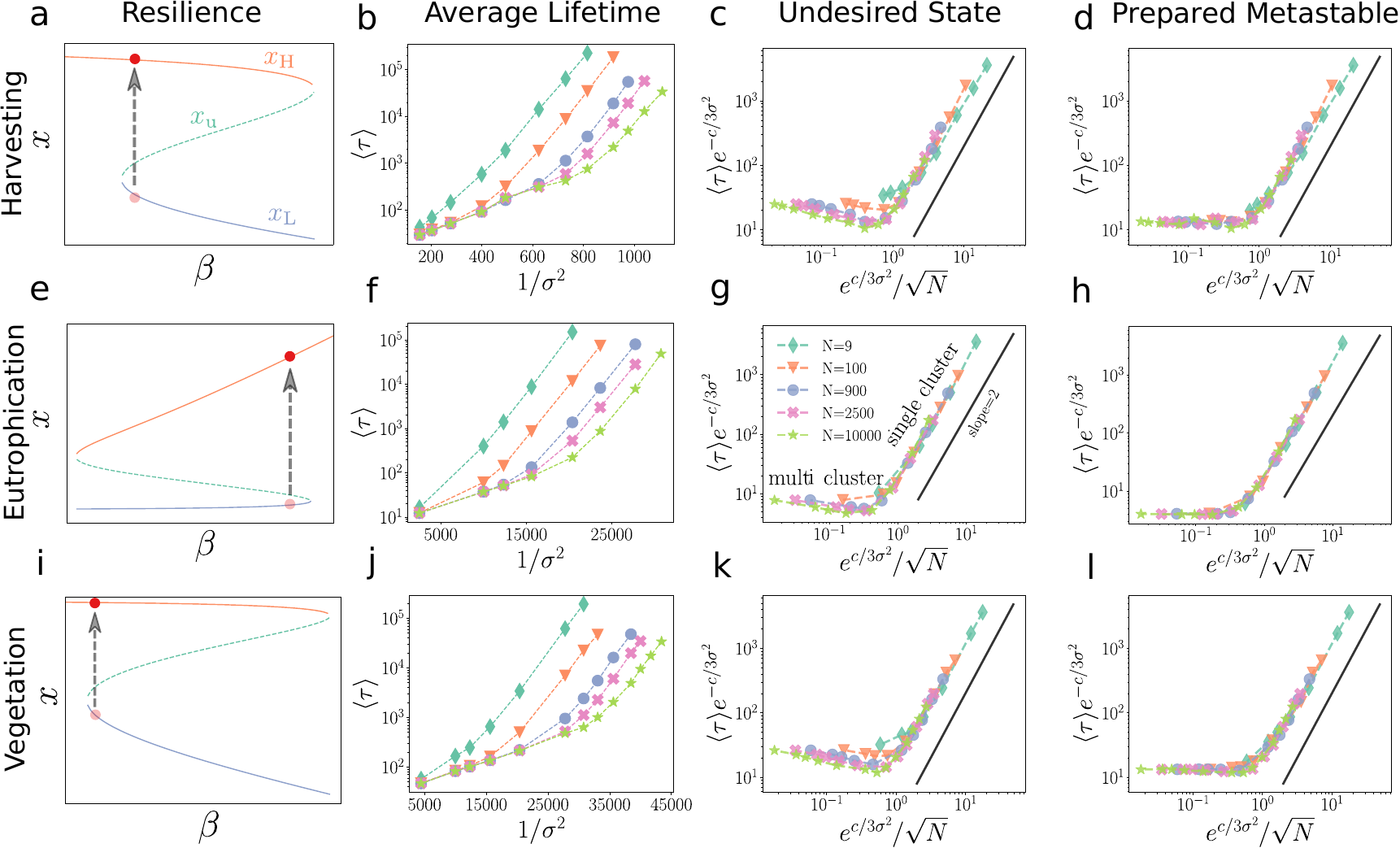}
    \caption{\textbf{Average lifetime $\langle \tau \rangle$ and the scaling results for three diffusion models.} The diffusion rate is set as $R= 0.02$. (a) -- (d) harvesting model, $r=1$ and $K=10$; (e) -- (h) eutrophication model, $a=0.5$ and $r=1$; (i) -- (l) vegetation model, $r=1$, $r_v=0.5$, $h_v=0.2$. (a), (e), and (i) are the resilience diagrams. These models exhibit alternative stale states for (a) $\beta \in (1.79, 2.60)$, (e) $\beta \in (0.86, 6.35)$, and (i) $\beta \in (2.59, 3.64)$. The arrows represent transitions from the low stable state to the high stable state. In simulation, the bifurcation parameter $\beta$ takes values of $1.80, 6.00, 2.60$ for (a), (e), and (i), respectively. For (b), (f), and (j), the system starts from the undesired states $x_{\mathrm{L}}$, and the average lifetime $\langle \tau \rangle $ changes with $N$ and $\sigma$. For (e), (g), and (k), the system starts from the undesired states $x_{\mathrm{L}}$, the lifetimes of two clusters and the crossover are shown in the scaling form. For (d), (h), and (l), the system starts from the prepared metastable states, the lifetimes of two clusters and the crossover are shown in the scaling fashion. }
    \label{fig:F7}
\end{figure}

As seen in Fig.~\ref{fig:F7}, the system size and noise strength decide the recovery time and the cluster mode.
Similarly, the average lifetime $\langle \tau \rangle $ of three diffusion models displays two distinct regimes. One is the single-cluster and the other is the multi-cluster regime (Fig.~\ref{fig:F7}a, \ref{fig:F7}d, \ref{fig:F7}g). The slope of $\ln\langle \tau \rangle$ versus $\sigma^{-2}$ reveals which mode is active. If the system starts from the undesired stable state, the scaling between two cluster modes deviates a little from the theoretical prediction (Fig.~\ref{fig:F7}b, \ref{fig:F7}e, \ref{fig:F7}h). If the system starts from the prepared state, resembling the metastable configuration, the scaling agrees well with the designed scaling function (Fig.~\ref{fig:F7}b, \ref{fig:F7}e, \ref{fig:F7}h).
Overall, the conclusions of the single-cluster and multi-cluster transition are validated in three diffusion models.

\section{\label{sec:discussion} Discussion}

We have utilized nucleation theory to analyze the noise-induced resilience restoration in ecosystems where the desired state has a much larger basin of attraction than the undesired state has. 
This is a general theory, and we successfully apply it to four ecological models, revealing the transition features.
During the restoration process, homogeneous nucleation theory distinguishes two different cluster modes: the single-cluster and multi-cluster transition modes.
We also derive the formulas for the recovery time under different conditions and propose a scaling function that collapses all the data onto one universal line.

The two cluster modes possess quite distinct features. The individual lifetime is random for the single-cluster phase, and the waiting time before the transition follows an exponential distribution. In contrast, for the multi-cluster mode, the lifetime is less random and centered about its average value so that the evolution of the global state $\rho$ is more deterministic. Which cluster mode the system follows is decided by its size and noise strength, and the crossover region is theoretically derived and can separate two phases. Generally, the large system subjected to strong noise presents the multi-cluster mode, and the small system with weak noise displays the single-cluster regime.
One quantity of interest for resilience restoration is the recovery time, which also depends on the system size and the noise strength. The rise of noise strength increases the nucleation rate, diminishing $\langle \tau \rangle $. There is no finite-size effect when noise is strong enough; that is, $\langle \tau \rangle $ is the same for different system sizes. The decrease of noise strength reveals the size effect, where larger systems take less time to complete transitions than smaller systems on average.  
Due to the distinct evolution features, the relationship between $\langle \tau \rangle $ and $\sigma$ varies for two cluster modes. What can be clearly seen in Fig.~\ref{fig:F3} and Fig.~\ref{fig:F7}a, \ref{fig:F7}d, \ref{fig:F7}g, is that $\langle \tau \rangle $ exhibits two distinctive regimes, corresponding to two cluster modes.
The scaling between two cluster modes is proposed according to Avrami's Law. The deviation observed in numerical simulation can be corrected by preparing the initial state close to metastable configurations to satisfy two homogeneous nucleation assumptions.

Employing the nucleation theory, we successfully extend the noise-induced transition in single-variable systems to spatially-extended multi-variable systems. Our framework is useful to predict critical transitions and guide the resilience restoration in the general dynamical systems presenting alternative stable states. 
One may wonder how the mean-field theory works in noisy environments as the noise-induced transition in single-variable systems has been well explored. The mean-field approach makes a satisfactory prediction of the system state when the system is close to the stable states. Unfortunately, it cannot capture the transition features observed in spatially-extended systems, including the average lifetime, the dependence on the system size, and the spatial-clustering patterns (see Supplementary Information, Sec II for the results of mean-field theory and the comparisons with the nucleation approach.).
Therefore, it is still an open question whether the mean-field theory can be used to study transitions in noisy environments. One may need to develop a new dimension reduction approach for this problem.
Also, there are further questions to be addressed. For example, we analyze the transition in the lattice model. In reality, the interaction relationship in ecosystems exhibits various structures. In addition, the interaction strength between components in the real-world complex systems varies, which may add to the difficulty of analysis. Furthermore, we focus on the one-way transition, which requires that the system be close to the theoretical bifurcation point where the basin of attraction of the undesired state vanishes. For the case when two alternative stable states have basins of attraction of similar sizes, stochastic switching (namely back-and-forth switching) between states may arise, which is beyond the scope of this study. In summary, network topology (in addition to spatial structure), coupling strength, and stochastic switching are of great interest to be investigated in future studies.

\section*{Acknowledgments}
We would like to thank three anonymous reviewers for their valuable and constructive comments and suggestions to improve the quality of our manuscript.
This work was supported by the Army Research Office (ARO) Grant W911NF-16-1-0524. The views and conclusions contained in this document are those of the authors and should not be interpreted as necessarily representing the official policies, either expressed or implied, of the U.S. Department of Defense.

\section*{Data availability statement}

The authors declare that data supporting the findings of this study are available within the paper.  

\section*{Code availability statement}

All codes for the reproduction of the reported results in this study are available upon request.

\section*{Author contributions}
C.M., G.K., B.K.S. and J.G. conceived the project and designed the research; C.M. implemented and performed numerical experiments and simulations;  C.M., G.K., B.K.S. and J.G. analyzed data and discussed results; C.M., G.K., B.K.S. and J.G. wrote, reviewed, and revised the manuscript.

\bibliographystyle{apsrev4-2_CN}

\bibliography{manuscript_v2}

\end{document}